\documentclass[11pt,twoside]{article}
\usepackage{asp2010}

\resetcounters

\begin{document}

\title{Binary properties of subdwarfs selected in the GALEX survey}
\author{Ad\'ela Kawka$^1$, Andrzej Pigulski$^2$, Simon O'Toole$^3$, 
St\'ephane Vennes$^1$, P\'eter N\'emeth$^1$, Andrew Williams$^4$, 
Lubomir Iliev$^5$, Zbyszek Ko{\l}aczkowski$^2$, Marek St\c{e}\'slicki$^2$}
\affil{$^1$Astronomick\'y \'ustav AV \v{C}R, Ond\v{r}ejov, Czech Republic}
\affil{$^2$Instytut Astronomiczny, Uniwersytet Wroc{\l}awski, Wroc{\l}aw, Poland}
\affil{$^3$Australian Astronomical Observatory, Epping, Australia}
\affil{$^4$Perth Observatory, Bickley, Australia}
\affil{$^5$Institute of Astronomy, BAS, Sofia, Bulgaria}

\begin{abstract}
We describe our programme to identify and analyse binary stars among
the bright subdwarfs selected in the Galaxy Evolution Explorer (GALEX) survey. 
Radial velocity time-series helped us identify 
subdwarfs with low-mass or compact stellar companions: We describe work
conducted on the bright binaries GALEX~J0321$+$4727
and GALEX~J2349$+$3844, and we present a radial velocity study of several objects
that include three new likely binaries.  We also carried out photometric observations 
that allowed us to detect long period pulsations in the subdwarf components in two 
of the close binaries.
\end{abstract}

\section{Introduction}

Hot subdwarf stars are found at the blue end of the horizontal branch and
are thought to evolve with a thin hydrogen envelope as a result of binary interaction. 
Approximately half of sdB stars are in close binary systems 
\citep[e.g.,][]{max2001,mor2003,gei2011}, either with a 
white dwarf or a cool main-sequence companion. This fraction of subdwarfs in 
close binaries can be explained by population synthesis studies such as those
of \citet{han2002,han2003}. Their models evolve binary stars through
either the common envelope or Roche-lobe overflow. On the other hand, single sdB stars 
could be the result of mergers of two low-mass helium white dwarfs.

Since the discovery of pulsations in sdB stars \citep{kil1997}, the internal
structure of sdB stars has become accessible to inquiries. Using asteroseismology the mass 
and hydrogen envelope thickness have been determined for a number of hot 
subdwarfs \citep[e.g.,][]{ran2007,cha2008}. Such studies have shown that
most sdB stars have masses between 0.4 and 0.5 M$_\odot$ and thin hydrogen
envelopes ($\log{M_{\rm H}/M_*} \sim -4$).

We have identified a sample of bright subdwarf candidates using ultraviolet
photometry from the GALEX all-sky survey and photographic magnitudes from the
Guide Star Catalog (GSC2.3.2). Details of the selection criteria are described
in \citet{ven2011}. 
We have now obtained high-quality spectroscopic observations for
$\sim$170  hot subdwarfs and conducted model atmosphere analyses. 
\citet{ven2011} presented the results of their analysis of 52 subdwarfs from this sample that
included measurements of their effective temperature, surface gravity and
helium abundance. An analysis of the complete sample
will be presented in N\'emeth, Kawka, \& Vennes (in preparation; see also these proceedings).

We have initiated a follow-up programme aimed at identifying new binary systems among
confirmed hot subdwarfs in the GALEX sample. Our first results on two
of these binaries, GALEX~J0321$+$4727 and GALEX~J2349$+$3844, were presented
in \citet{kaw2010}. Here we report our progress in identifying and analysing
the properties of subdwarfs in close binaries.

\section{Observations}

We observed several bright GALEX targets using the spectrograph at the coud\'e focus of
the 2 m telescope at Ond\v{r}ejov. We used the 830.77 lines/mm grating with a
SITe $2030\times 800$ CCD providing a resolution of $R=13\,000$ and a spectral
range from 6254 \AA\ to 6763 \AA\ \citep{sle2002}. We also obtained spectra 
centred on H$\alpha$ with the coud\'e spectrograph attached to the 2 m 
telescope at Rozhen Observatory. We used the 632 lines/mm grating with a SITe 
$1024\times1024$ CCD providing a resolution of $R=28\,000$.

We have also started a similar programme in the southern hemisphere
using the Wide Field Spectrograph \citep[WiFeS,][]{dop2007} attached to the 
2.3 m telescope at Siding Spring Observatory (SSO). The first set of data
was obtained between UT 2010 July 14 and 18. We used the B3000 and R7000
gratings in the blue and red, respectively.

We have obtained follow-up photometry for several of our targets with the 0.6 m
telescope at Bia{\l}k\'ow Observatory (7 objects) and 0.6 m telescope at Perth 
Observatory (46 objects). Here, we present a summary of observations of 
GALEX~J0321$+$4727 and GALEX~J2349$+$3844 that were obtained at Bia{\l}k\'ow 
Observatory. GALEX~J0321$+$4727 was observed during 6 nights between UT 2010 
October 9 and November 14 and GALEX~J2349$+$3844 was observed during 6 nights 
between UT 2010 October 9 and December 4.

\begin{figure}[t]
\setlength{\textwidth}{10.0cm}
\plotone{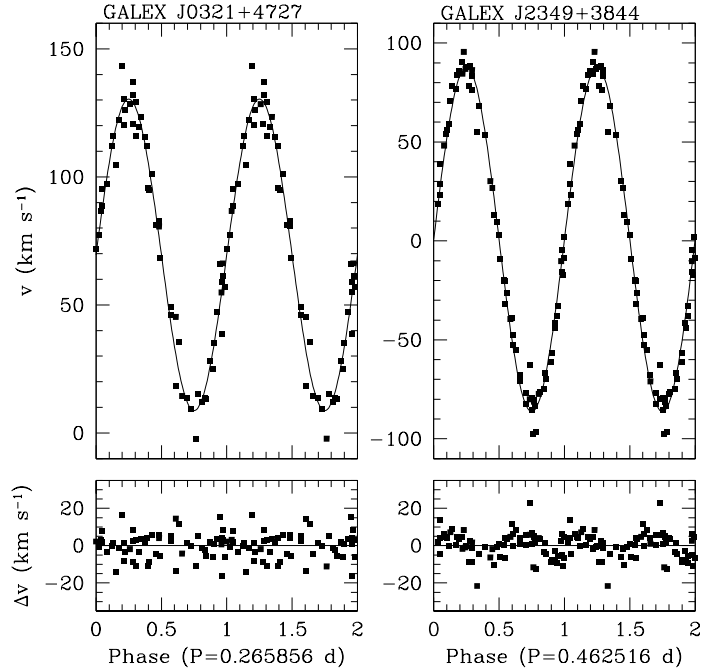}
\caption{Radial velocity measurements of GALEX~J0321$+$4727 and GALEX~J2349$+$3844 phased on the
orbital period. Radial velocity residuals in GALEX~J2349$+$3844
reveal the eccentricity of the orbit. \label{fig_vel}}
\end{figure}

\section{Binary Parameters}

We determined the radial velocities by measuring the centre of the H$\alpha$ 
core. The velocities were adjusted to the solar system barycentre. 
Table~\ref{tbl-bin} updates the orbital parameters for GALEX~J0321$+$4727 and 
GALEX~J2349$+$3844 that were originally presented in \citet{kaw2010}.
Figure~\ref{fig_vel} shows the new radial velocity curves and residuals to the best-fitting sinusoidal curve. 
The residuals in the radial velocity curve of 
GALEX~J2349$+$3844 show systematic deviations revealing
an orbital eccentricity $e = 0.06\pm0.02$. The short orbital period
of the binary implies that it must have evolved through at least one common 
envelope (CE) phase \citep{han2002}. The CE phase is expected to circularize 
the orbit and therefore a measurable eccentricity is not expected for 
post-CE binaries. \citet{ede2005} reported similar periodic residual patterns 
for a small number of close post-CE binaries containing a subdwarf also
interpreted as orbital eccentricity. The eccentricity may be the result of
past interaction with a third body or anisotropic mass-loss.

\begin{table}[t!]
\caption{Binary parameters\label{tbl-bin}}
\centering
\begin{tabular}{lcc}
\tableline
\tableline\\
Parameter & GALEX~J0321$+$4727 & GALEX~J2349$+$3844 \\
\tableline\\
Period (d)                   & $0.265856\pm0.000003$ & $0.462516\pm0.000005$ \\
$T_0$ (BJD 2455000+)         & $45.583\pm0.003$      & $45.510\pm0.002$     \\
K (km~s$^{-1}$)              & $60.8\pm4.5$          & $87.9\pm2.2$        \\
$\gamma$ (km~s$^{-1}$)       & $69.6\pm2.2$          & $2.0\pm1.0$         \\
$f(M_{\rm sec})$ ($M_\odot$) & $0.00619\pm0.00009$   & $0.03167\pm0.00033$ \\
$e$                          &   (0)                 &  $0.06\pm0.02$      \\
$T_{\rm eff}$ (K)            & $27990^{+460}_{-400}$ & $23770^{+330}_{-350}$ \\
$\log{g}$                    & $5.34\pm0.07$         & $5.38^{+0.05}_{-0.06}$ \\
$\log{({\rm He}/{\rm H})}$   & $-2.52^{+0.17}_{-0.22}$ & $-3.44^{+0.25}_{-0.30}$ \\\\
\tableline
\end{tabular}
\end{table}

We determined the atmospheric parameters of the two subdwarfs by fitting them
with synthetic spectra that were calculated using TLUSTY/SYNSPEC codes.
For details of the fitting see N\'emeth, Kawka \& Vennes (these proceedings).
The temperature and surface gravity for GALEX~J2349$+$3844 have been remeasured using the complete
Balmer line series. The temperature is 4700 K cooler and the gravity is 0.5 dex
lower than originally estimated by 
\citet{kaw2010} who relied on H$\alpha$ only.

We have obtained spectroscopic observations of several other bright GALEX subdwarfs
using the Ond\v{r}ejov 2 m telescope and the 2.3 m telescope at SSO. Most of the 
objects do not show significant velocity variations, however there are a few
interesting cases that deserve further observations. GALEX~J2038$-$2657 
is a hot sdO star with a G type subgiant (III-IV) (N\'emeth et al. in 
preparation). We obtained two spectra, a day apart, that show variable
H$\alpha$ emission.
Table~\ref{tbl-rad} 
lists the number of spectra obtained for each star, as well as the average and dispersion
of the velocity measurements. This sample also includes two subdwarf candidates 
(TYC4000-216-1 and TYC4406-285-1) from \citet{jim2011}. We added to our sample the ``constant''
velocity star Feige~66 and adopted its velocity dispersion as representative of 
1$\sigma$ measurement uncertainty at Ond\v{r}ejov.

\begin{table}
\caption{Summary of radial velocities for a sample of GALEX subdwarfs\label{tbl-rad}}
\centering
\begin{tabular}{lcccc}
\tableline
\tableline\\
Name          & Site & N  & $<v>$ (km~s$^{-1}$) & $\sigma_v$ (km~s$^{-1}$) \\
\tableline\\
GALEX~J1526$+$7941  & Ond\v{r}ejov &  5 &   5.2 & 25.0 \\
GALEX~J0639$+$5156  & Ond\v{r}ejov &  8 &  98.9 & 22.0 \\
GALEX~J1348$+$4337  & Ond\v{r}ejov &  6 & -25.0 & 14.0 \\
GALEX~J1632$+$8513  & Ond\v{r}ejov &  2 & -47.9 &  9.1 \\
GALEX~J0747$+$6225  & Ond\v{r}ejov &  8 & -76.4 &  8.8 \\
GALEX~J1702$+$6353  & Ond\v{r}ejov &  4 & -12.9 &  8.2 \\
GALEX~J1600$-$6433  & SSO          &  7 &  53.8 &  7.3 \\
TYC4000-216-1       & Ond\v{r}ejov & 18 &  36.3 &  7.1 \\
GALEX~J1902$-$5130  & SSO          &  5 &-111.0 &  5.2 \\
GALEX~J1911$-$1406  & SSO          &  3 &-151.9 &  4.9 \\
TYC4406-285-1       & Ond\v{r}ejov & 13 & -24.7 &  4.6 \\
GALEX~J1753$-$5007  & SSO          &  4 & -59.8 &  4.3 \\
GALEX~J1736$+$2806  & Ond\v{r}ejov & 21 & -11.8 &  4.1 \\
Feige 66            & Ond\v{r}ejov & 10 &  -3.6 &  3.9 \\
GALEX~J1017$+$5516  & Ond\v{r}ejov &  4 &  73.6 &  3.7 \\
GALEX~J1632$+$0759  & SSO          &  2 & -52.4 &  3.3 \\
GALEX~J2153$-$7003  & SSO          &  5 &  43.0 &  1.2 \\
\tableline
\end{tabular}
\end{table}

Three objects show a dispersion in velocity measurements larger than 3$\sigma$ including 
GALEX~J0639$+$5156, which is also a V361 Hya pulsating subdwarf (Vu\v{c}kovi\'c 
et al. these proceedings). The most promising object that shows variable 
velocity measurements is GALEX~J1526$+$7941. Although we have only obtained five exposures so far,
the difference in velocity between two consecutive nights was $\sim$60 
km~s$^{-1}$. GALEX~J1911$-$1406 is a hot sdO star; we used 
HeII$\lambda$ 6560.09\AA\ line to measure the radial velocities which do not appear to vary. 

GALEX~J1736$+$2806 is a composite sdB plus main-sequence G system 
\citep{ven2011}, and it does not show variability. This is one of the objects
for which we obtained a reasonably large number of spectra over a period of 
two years. Therefore, either the binary is at a low-inclination, or more 
likely the system passed through a Roche-lobe overflow phase resulting in a 
wide binary with a long period $\sim$1000 days \citep{han2002}.

\section{Photometric variations}

\begin{figure}[t]
\setlength{\textwidth}{10.0cm}
\plotone{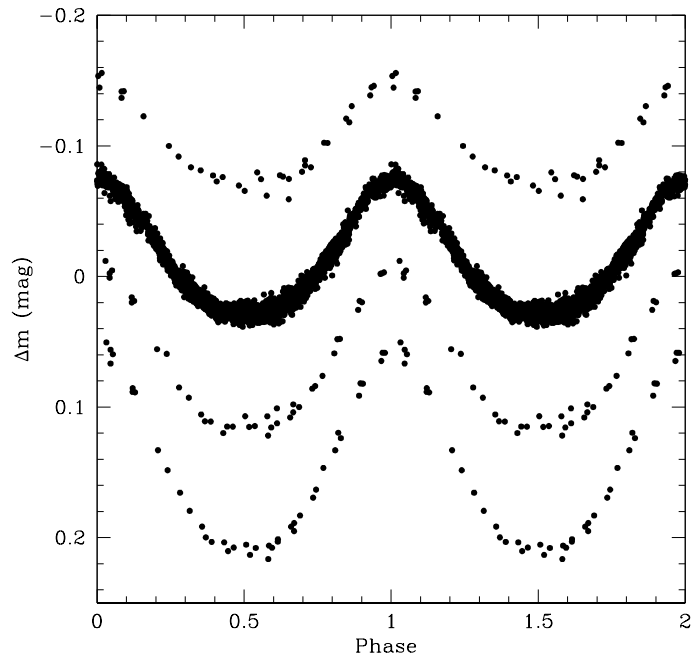}
\caption{The photometry of GALEX~J0321$+$4727 is phased on the orbital
period determined from radial velocity measurements and shows the reflection 
effect. The photometric bands plotted from top to bottom are $B$, $V$, $R$ and
$I$. \label{fig_phot_ref}}
\end{figure}

\begin{figure}[t]
\setlength{\textwidth}{11.0cm}
\plotone{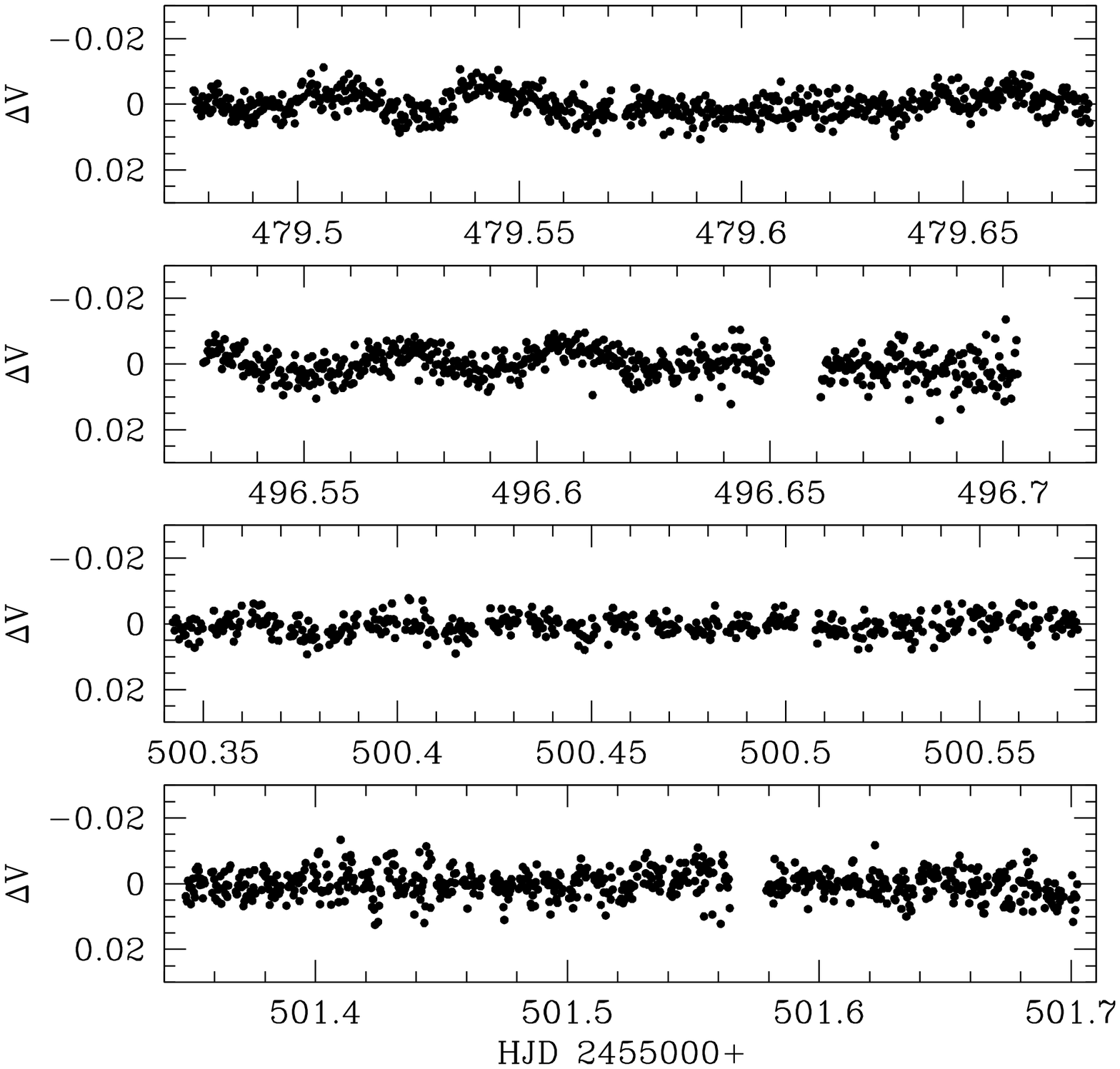}
\plotone{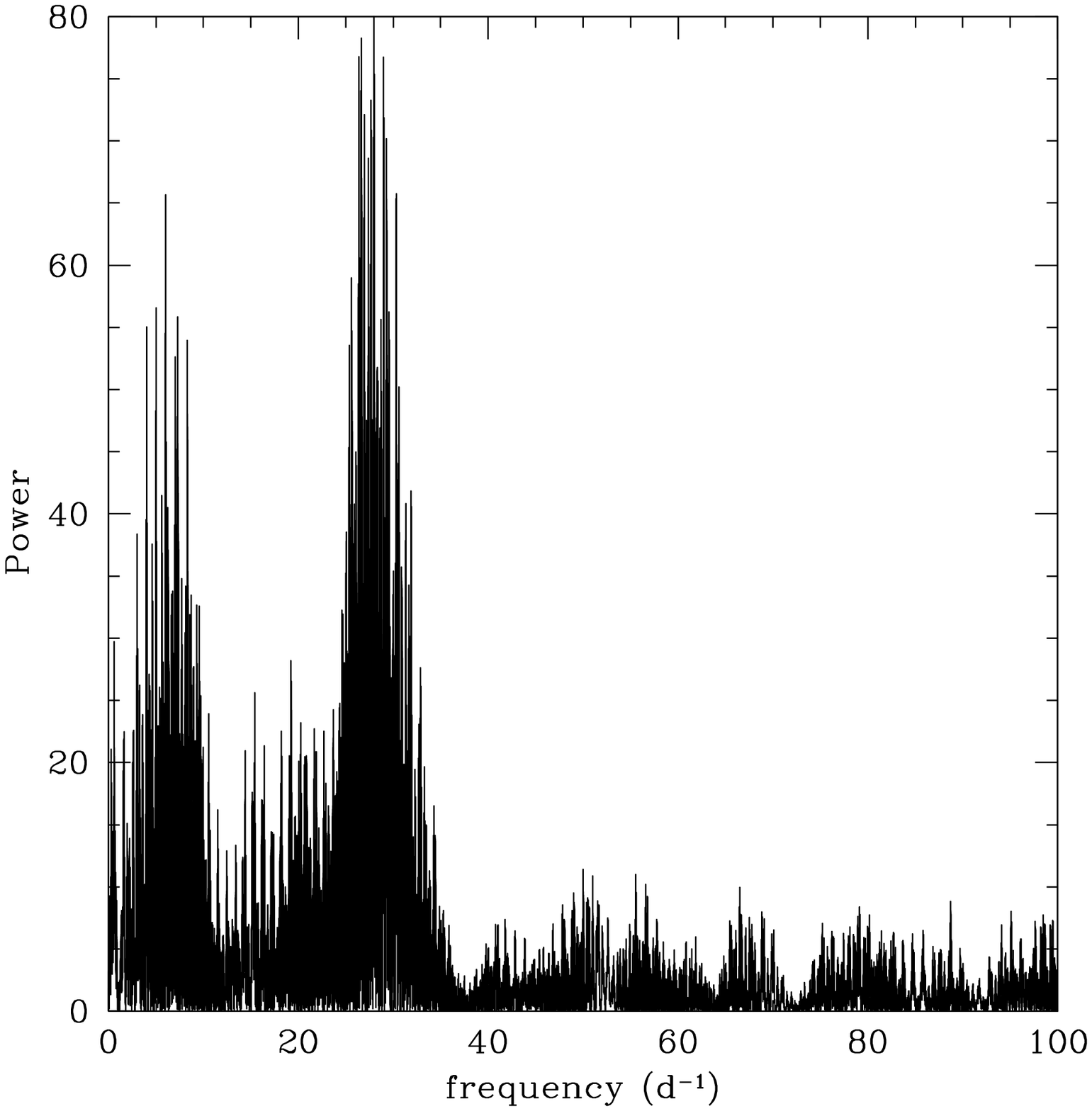}
\caption{Photometry of GALEX~J0321$+$4727 ({\it top}) with the reflection
effect removed and the power spectrum ({\it bottom}) showing the main pulsation 
frequencies.
\label{fig_phot_gal0321}}
\end{figure}

Based on an analysis of NSVS light curves
GALEX~J0321$+$4727 was reported to be photometrically variable
\citep{kaw2010}. The NSVS photometric measurements
have relatively large uncertainties and were obtained sporadically, and, therefore,
timeseries photometry seems preferable.
Photometric time series of GALEX~J0321$+$4727 were obtained in $BVRI$.
In allowing us to accurately phase the data with our new ephemeris, these
time series confirmed the reflection effect of the subdwarf by the 
cool companion. Figure~\ref{fig_phot_ref} shows the photometric measurements of
GALEX~J0321$+$4727 in the four bands folded on the best orbital period. As
expected for the reflection effect, the highest amplitude variations are 
observed in the $I$ band ($\Delta I \sim 0.08$ mag) and the lowest in the $B$ 
band ($\Delta B \sim 0.04$ mag).
The photometric observations also revealed weaker variations indicating that
GALEX~J0321$+$4727 is a pulsating subdwarf. Figure~\ref{fig_phot_gal0321} 
shows the pulsation lightcurve after removing the variations due to the
reflection effect. We were able to detect two
frequencies in our data:  $\sim$28.0 d$^{-1}$ and $\sim$6.0 
d$^{-1}$ with the respective amplitudes of $\sim$1.3 mmag and 
$\sim$1.4 mmag.

\begin{figure}[t]
\setlength{\textwidth}{11.0cm}
\plotone{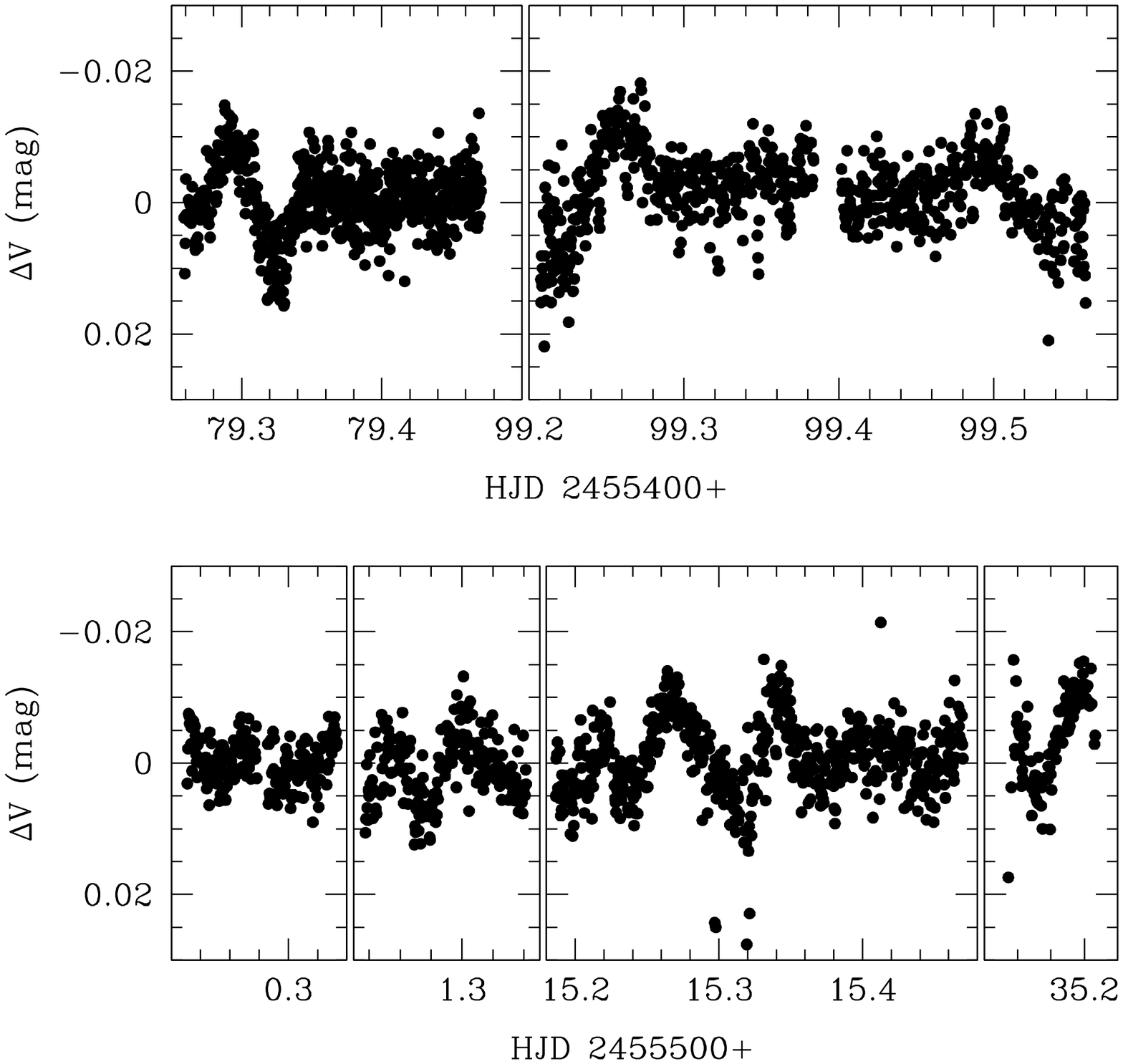}
\plotone{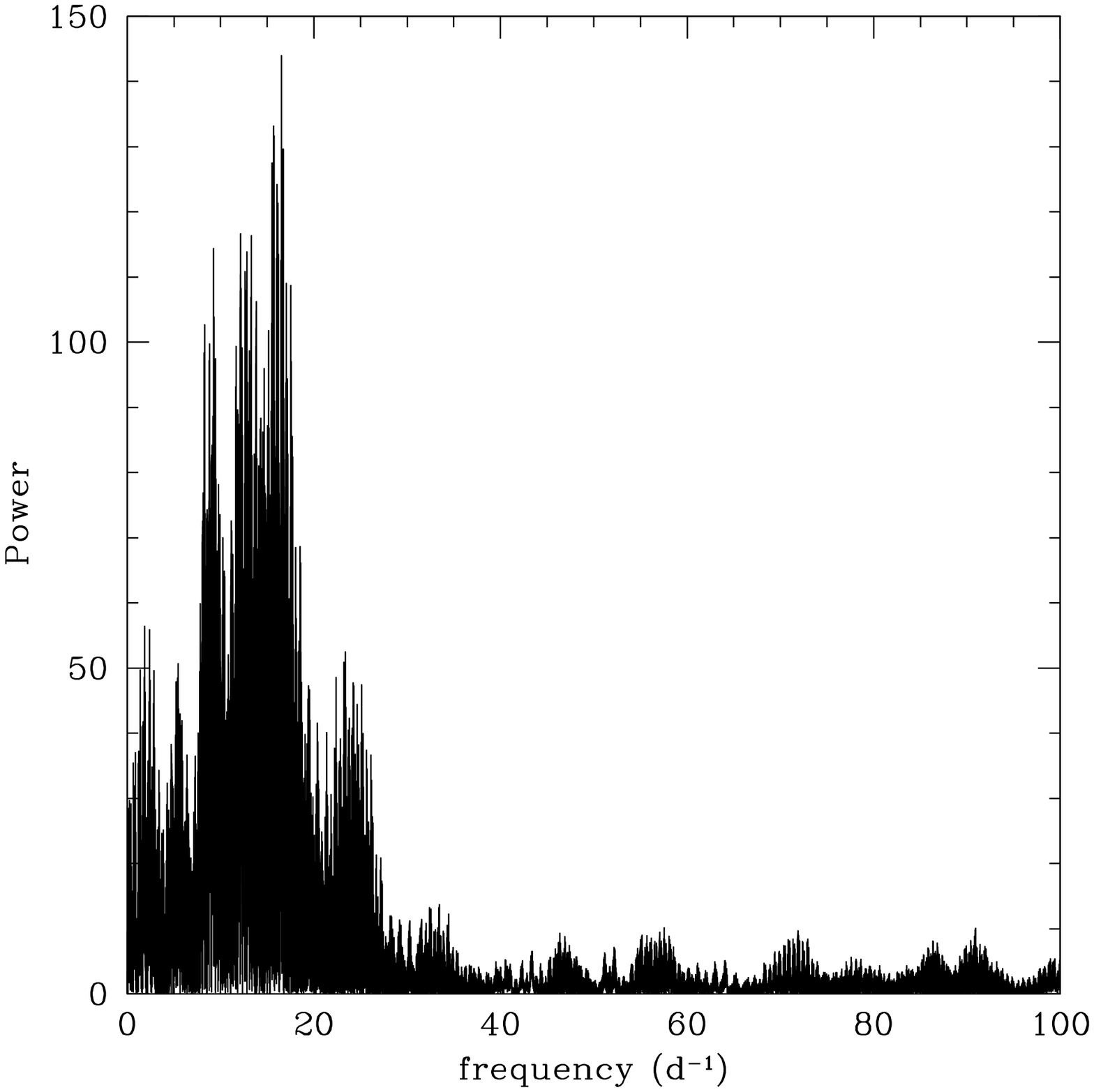}
\caption{Photometry of GALEX~J2349$+$3844 ({\it top}) and the power
spectrum ({\it bottom}) showing the main pulsation frequencies.
\label{fig_phot_gal2349}}
\end{figure}

Figure~\ref{fig_phot_gal2349} shows the photometric $V$ measurements of 
GALEX~J2349$+$3844 revealing it to be a pulsating subdwarf. We detected three
frequencies in our data: $\sim$16.5 d$^{-1}$, $\sim$12.2 
d$^{-1}$ and $\sim$9.1 d$^{-1}$ with the respective amplitudes of
$\sim$2.9, $\sim$2.8 and $\sim$2.2 mmag. The low frequencies 
place GALEX~J0321$+$4727 and GALEX~J2349$+$3844 among the group of slowly 
pulsating V2093 Her type subdwarfs 
\citep{gre2003}. GALEX~J2349$+$3844 is at the red edge of the instability
strip of V2093 Her pulsators, while GALEX~J0321$+$4727 is closer to the
blue edge.

\section{Summary}

We are currently conducting a radial velocity and photometric survey of 
the brightest hot subdwarf stars selected
in the GALEX survey. With only two previously confirmed close binary systems, 
GALEX~J0321$+$4727 and GALEX~J2349$+$3844, our sample appears to have only 
a $\sim$10\% success rate of detection. However, with the inclusion of the three 
potentially variable stars, this rate increases to $\sim$30\%. 
The aims of the photometric survey are to identify new candidates for 
pulsation studies and for binarity via the detection of the reflection effect. 
We report the detection of pulsation periods of $\sim$1-1.5 hrs in
the subdwarf stars GALEX~J0321$+$4727 and GALEX~J2349$+$3844. Therefore, the two
stars belong to the V2093~Her class of pulsators \citep{gre2003}.

\acknowledgements A.K., P.N., and S.V. are supported by GA AV grant numbers IAA300030908 
and IAA301630901, and by GA \v{C}R grant number P209/10/0967.
A.P. acknowledges Polish MNiSzW grant N N203 302635. We thank Dawid 
Mo\'zdzierski for sharing a night with us.

\bibliography{kawka}

\end{document}